\documentclass[%
 aip,
 jmp,%
 amsmath,amssymb,
 preprint,%
]{revtex4-1}

\usepackage{graphicx,xcolor}
\usepackage{dcolumn}
\usepackage{bm}
\usepackage{subfig}
\usepackage{float}
\usepackage{caption}

\begin{document}

\preprint{AIP/123-QED}

\title[]{Cooling Dynamics Through Transition Temperature of Niobium SRF Cavities Captured by Temperature Mapping}

\author{M. Martinello}
 \email{mmartine@fnal.gov}
\affiliation{Fermi National Accelerator Laboratory, Batavia, Illinois 60510, USA.}%
\affiliation{Department of Physics, Illinois Institute of Technology, Chicago, Illinois 60616, USA.}%
\author{A. Romanenko}
 \email{aroman@fnal.gov}
\affiliation{Fermi National Accelerator Laboratory, Batavia, Illinois 60510, USA.}%
\author{M. Checchin}
\affiliation{Fermi National Accelerator Laboratory, Batavia, Illinois 60510, USA.}%
\affiliation{Department of Physics, Illinois Institute of Technology, Chicago, Illinois 60616, USA.}%
\author{A . Grassellino}
\affiliation{Fermi National Accelerator Laboratory, Batavia, Illinois 60510, USA.}%
\author{A. C. Crawford}
\affiliation{Fermi National Accelerator Laboratory, Batavia, Illinois 60510, USA.}%
\author{A. Melnychuk}
\affiliation{Fermi National Accelerator Laboratory, Batavia, Illinois 60510, USA.}%
\author{D. A. Sergatskov}
\affiliation{Fermi National Accelerator Laboratory, Batavia, Illinois 60510, USA.}%
\date{\today}

\begin{abstract}
Cool-down dynamics of superconducting accelerating cavities became particularly important for obtaining very high quality factors in SRF cavities. Previous studies proved that when cavity is cooled fast, the quality factor is higher than when cavity is cooled slowly. This has been discovered to derive from the fact that a fast cool-down allows better magnetic field expulsion during the superconducting transition. In this paper we describe the first experiment where the temperature all around the cavity was mapped during the cavity cool-down through transition temperature, proving the existence of two different transition dynamics: a sharp superconducting-normal conducting transition during fast cool-down which favors flux expulsion and nucleation phase transition during slow cool-down, which leads to full flux trapping.
\end{abstract}

\maketitle
\indent It has been shown \cite{1} that ambient magnetic field can be either fully trapped or fully expelled during the superconducting (SC) transition depending on the details of cavity cool-down through transition temperature.\\
\indent In particular it has been shown that when the cool-down occurs fast, high thermogradients are present along the cavity; in this case it has been suggested that the superconducting-normal conducting (SC-NC) interface moves sharply promoting efficient magnetic field expulsion. On the other hand, when the cool-down occurs slowly, the thermogradients along the cavity are minimized and the SC-NC transition does not appear sharp but rather the nucleation regime involves the creation of normal conducting islands surrounded by superconducting phase, from which magnetic flux expulsion becomes less favorable, and therefore promoting magnetic flux trapping into these regions leading to high surface resistance \cite{2,3}.\\
\indent In this paper we present for the first time the temperature mapping \cite{4} of the cavity during a fast and a slow cool-down, in order to visualize how the superconducting transition occurs in the two cases, and to verify the veridicity of previous hypothesis.\\
\indent In this experiment the cavity was placed horizontal respect to the cooling flux, at the Fermilab Vertical Test facility (VTS), in order to study the cavity behavior in the same condition as cavity into the cryomodule of an accelerator. Temperature sensors were fixed on $36$ boards, each equipped with $16$ thermometers, in order to monitor the temperature all around the cavity. The thermometer number $8$ of board $31$ is located on the bottom of the cavity, whereas the thermometer number $8$ of board $13$ is located on top of the cavity.\\
\indent The fast cool-down temperature mapping refers to a cool-down with starting temperature of $250 K$, the starting temperature of the slow cool-down is $12 K$ instead. In the case of slow cool-down the starting temperature is close to the critical temperature in order to maintain the temperature as uniform as possible along the cavity during the transition.\\
\indent The fast cool-down T-map images are shown in Fig. \ref{Fast}, while the slow cool-down images are shown in Fig. \ref{Slow}. In either cases the white color indicates the superconducting phase, which means the temperature measured is lower than $T_{c}=9.25K$.
%
%
\captionsetup[figure]{labelfont={color=white}}
\begin{figure}[H]%
\centering
\subfloat[][]{\includegraphics[]{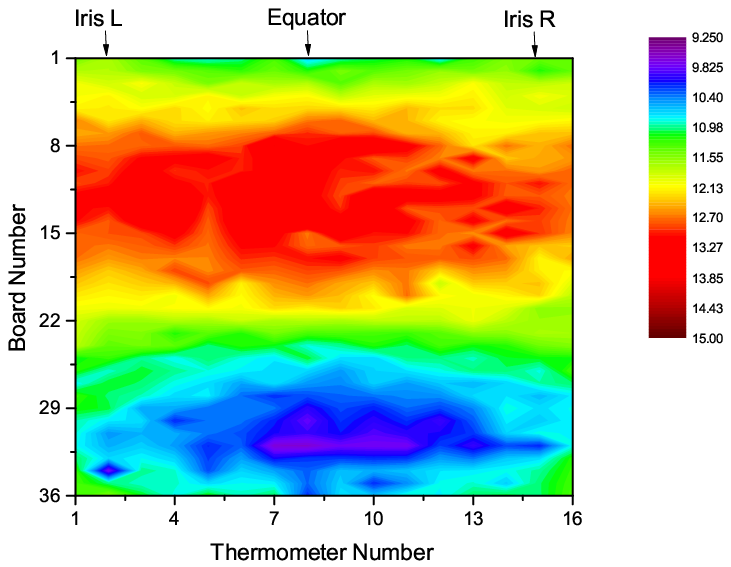}}%
\subfloat[][]{\includegraphics[]{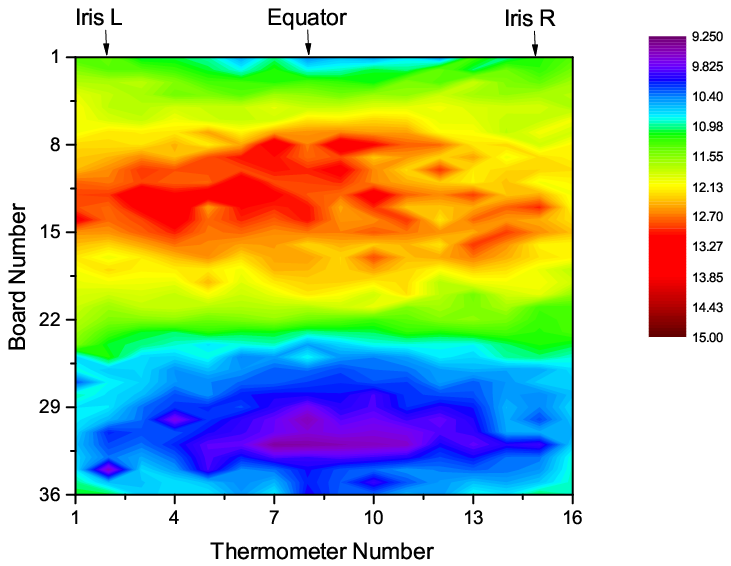}}\\
\subfloat[][]{\includegraphics[]{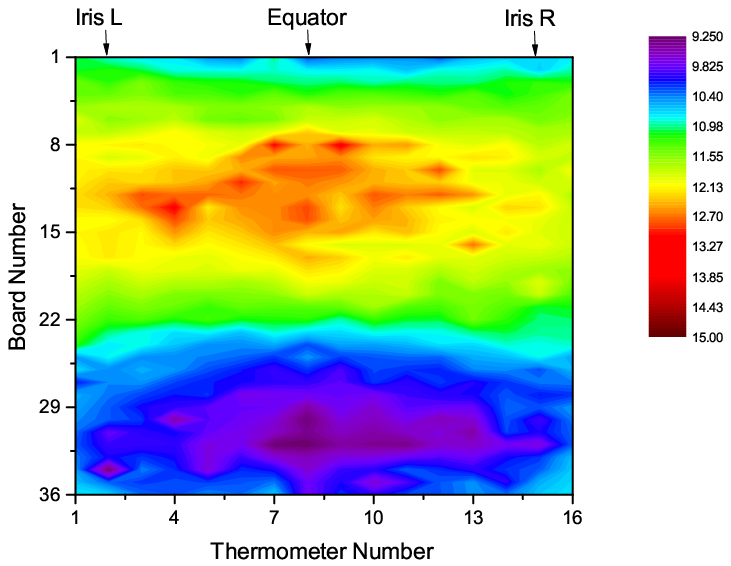}}%
\subfloat[][]{\includegraphics[]{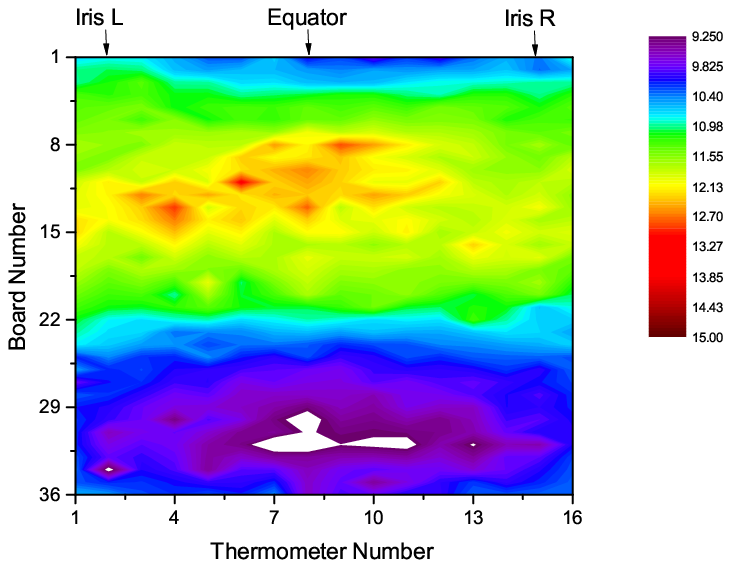}}\\
\subfloat[][]{\includegraphics[]{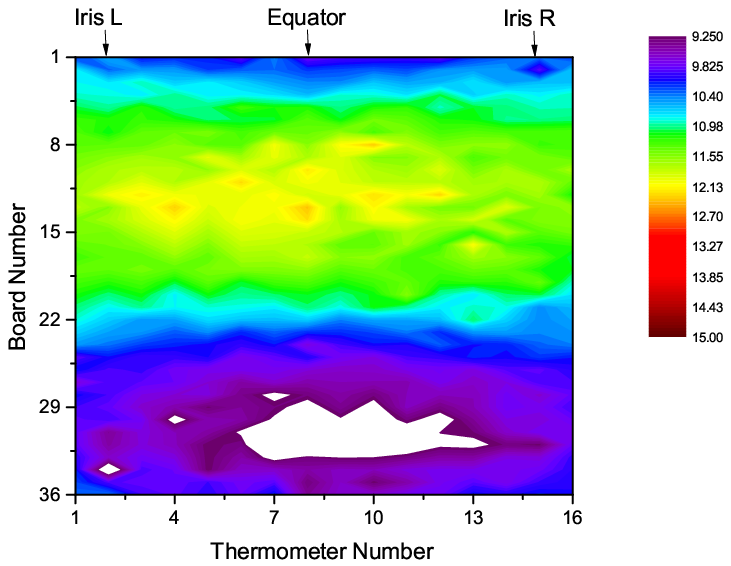}}%
\subfloat[][]{\includegraphics[]{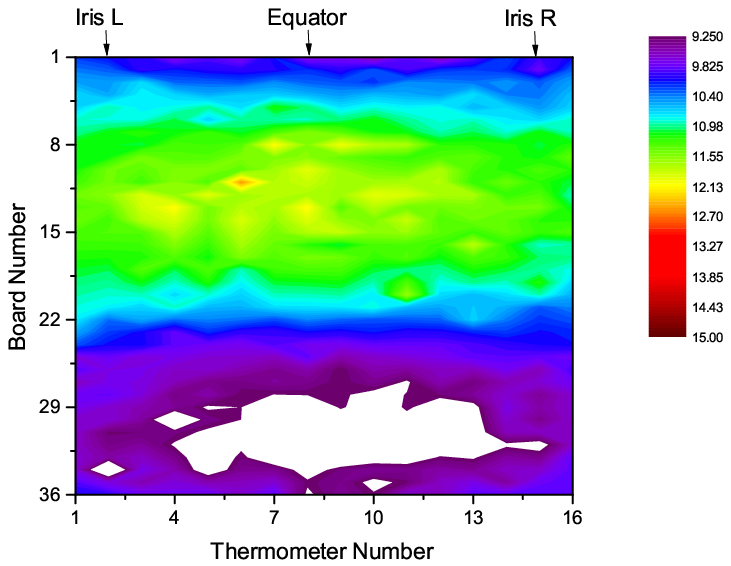}}\\
\caption{}%
\end{figure}
\begin{figure}[H]%
\ContinuedFloat
\centering
\subfloat[][]{\includegraphics[]{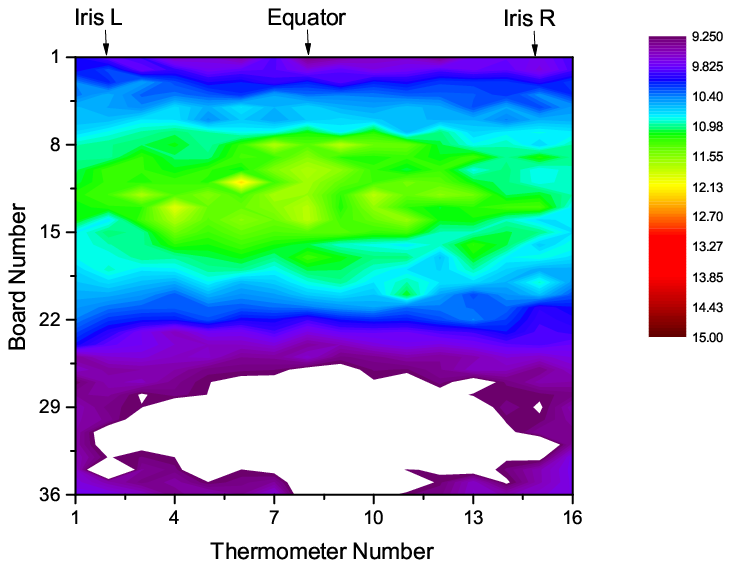}}%
\subfloat[][]{\includegraphics[]{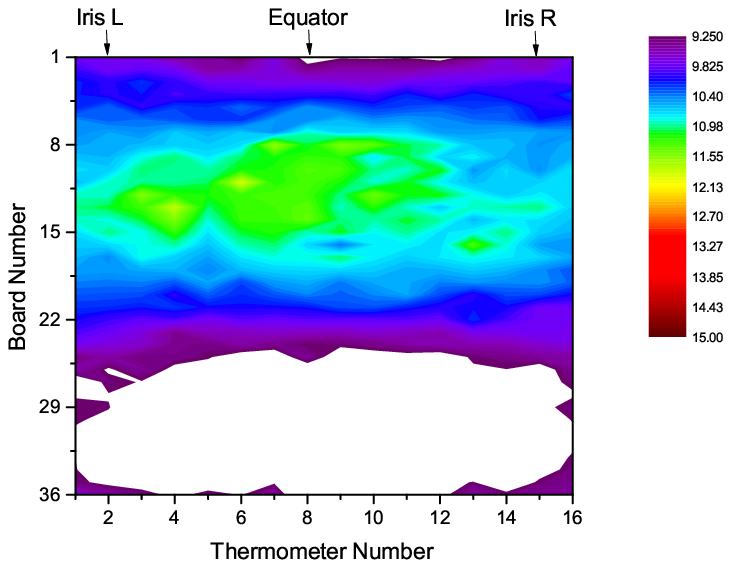}}\\
\subfloat[][]{\includegraphics[]{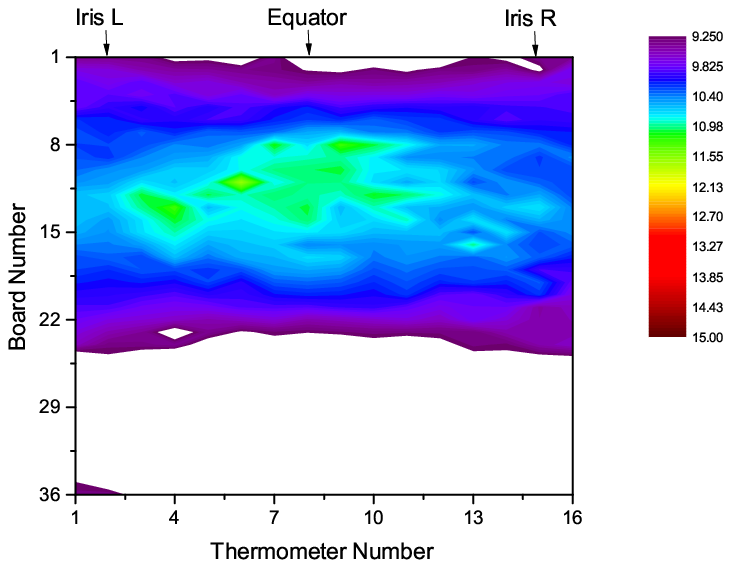}}%
\subfloat[][]{\includegraphics[]{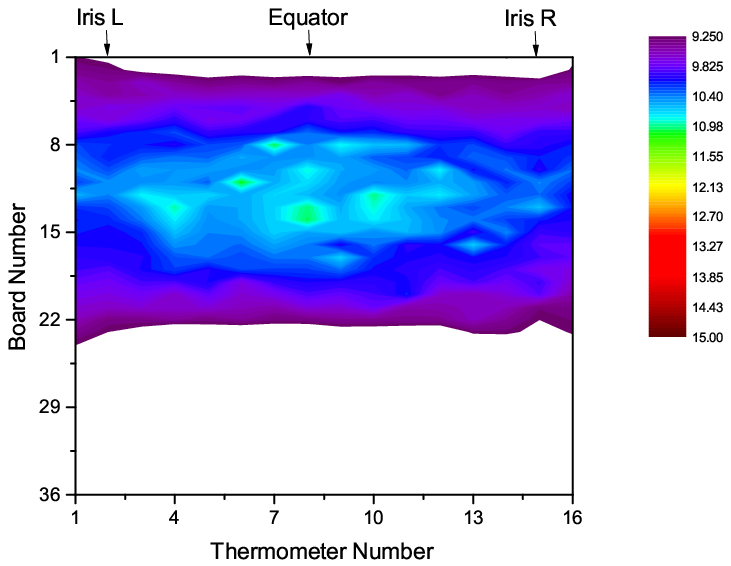}}\\
\subfloat[][]{\includegraphics[]{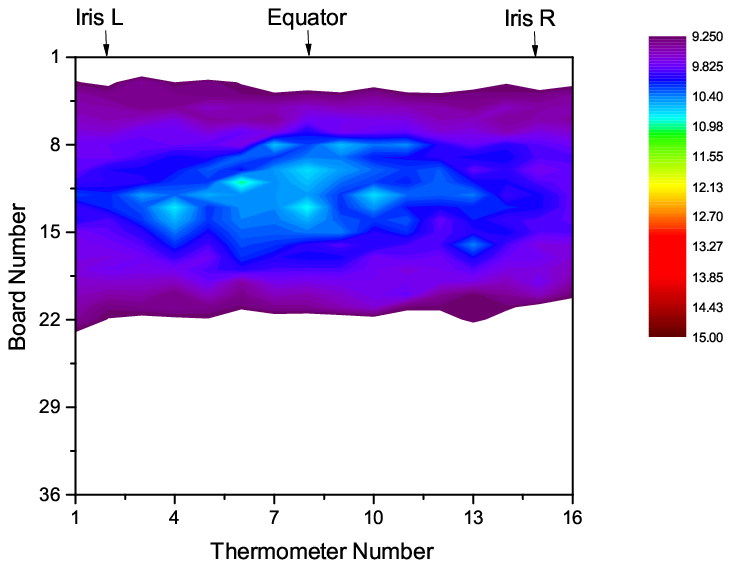}}%
\subfloat[][]{\includegraphics[]{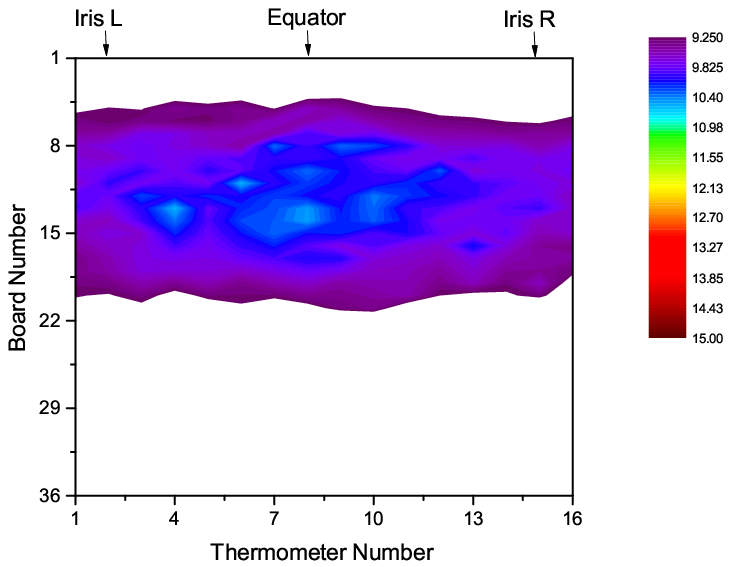}}\\
\caption{}%
\end{figure}
\captionsetup[figure]{labelfont={color=black}}
\begin{figure}[H]%
\ContinuedFloat
\centering
\subfloat[][]{\includegraphics[]{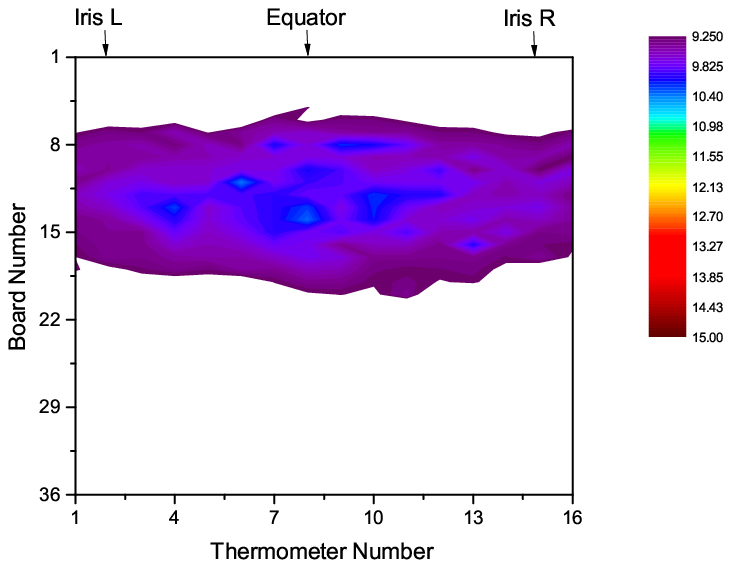}}%
\subfloat[][]{\includegraphics[]{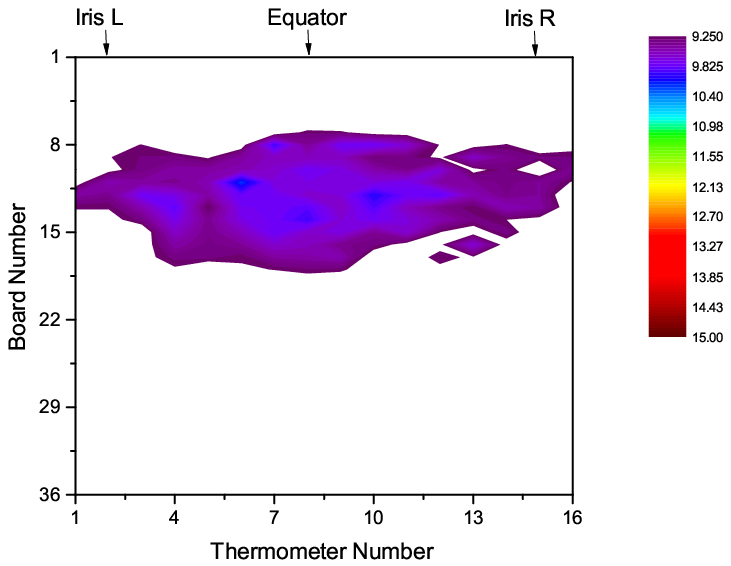}}\\
\subfloat[][]{\includegraphics[]{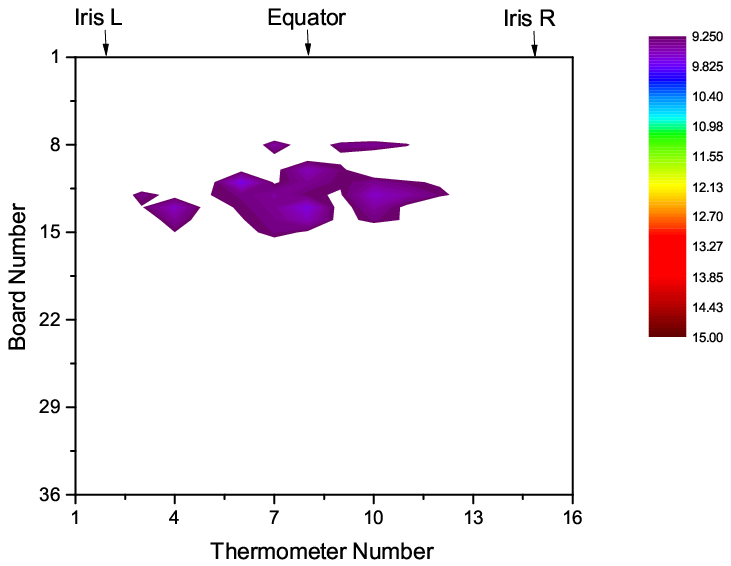}}%
\subfloat[][]{\includegraphics[]{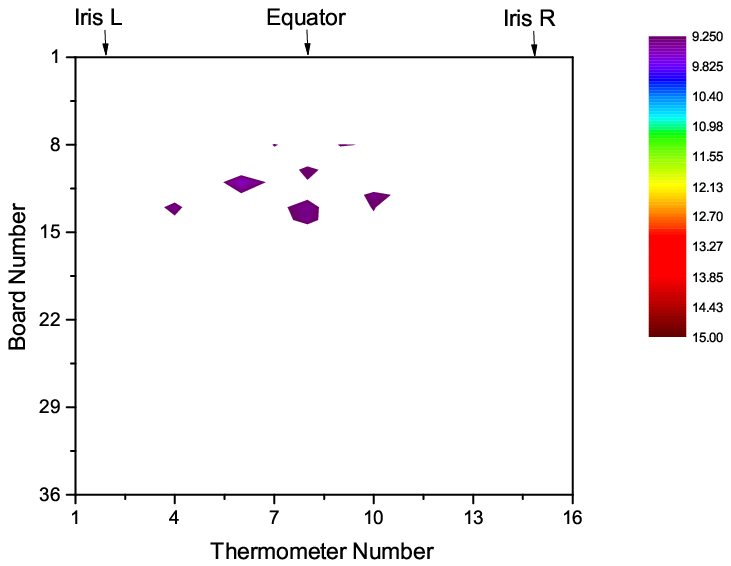}}\\
\subfloat[][]{\includegraphics[]{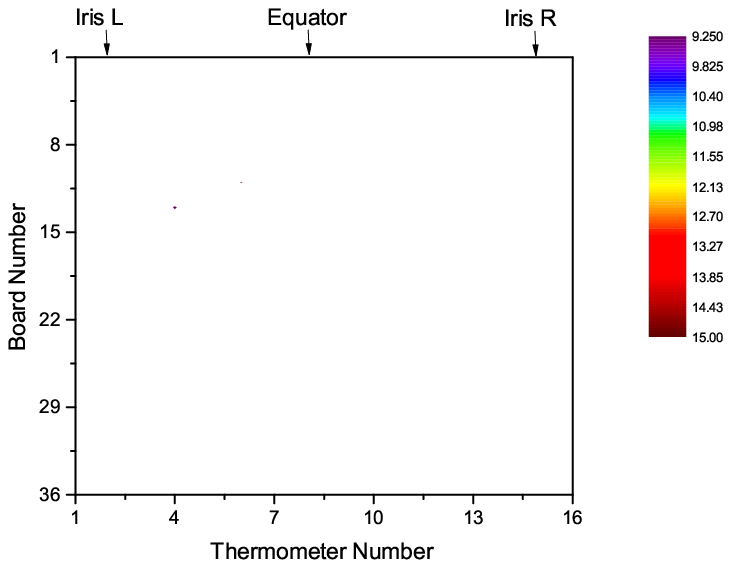}}%
\subfloat[][]{\includegraphics[]{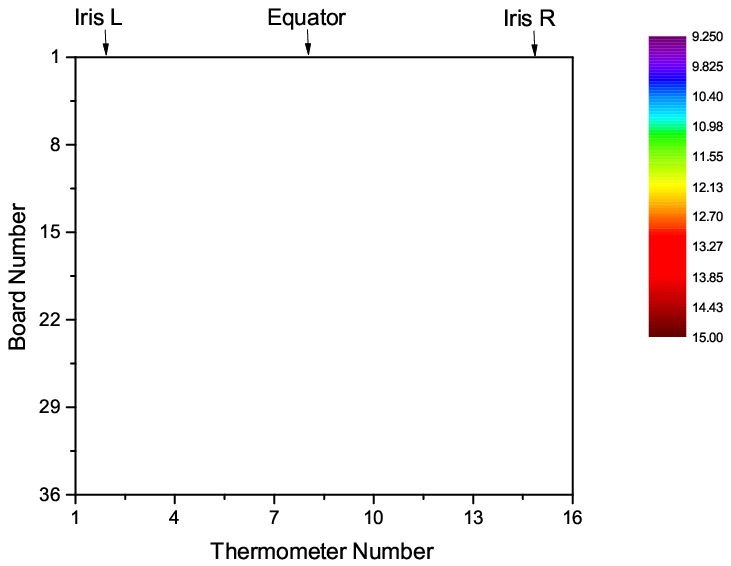}}\\
\caption{Chronological sequence of T-map images acquired during a fast cool-down from 250 K.}%
\label{Fast}
\end{figure}
\clearpage
%
%
\captionsetup[figure]{labelfont={color=white}}
\begin{figure}[H]%
\centering
\subfloat[][]{\includegraphics[]{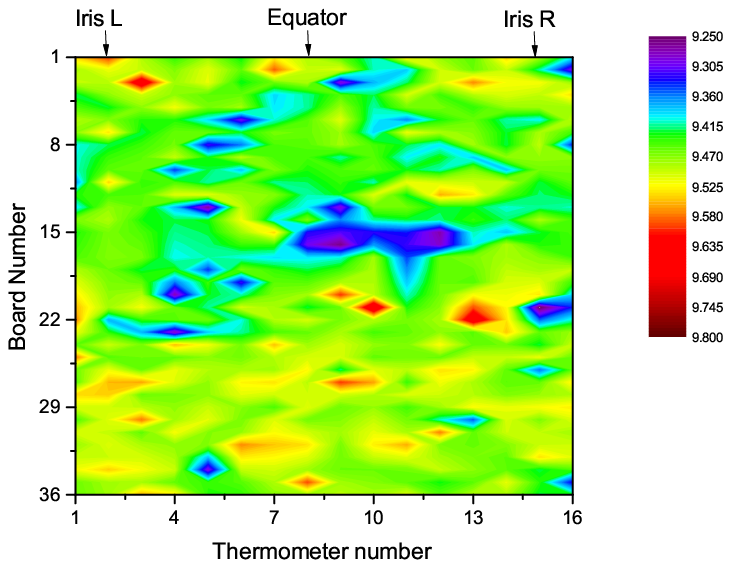}}%
\subfloat[][]{\includegraphics[]{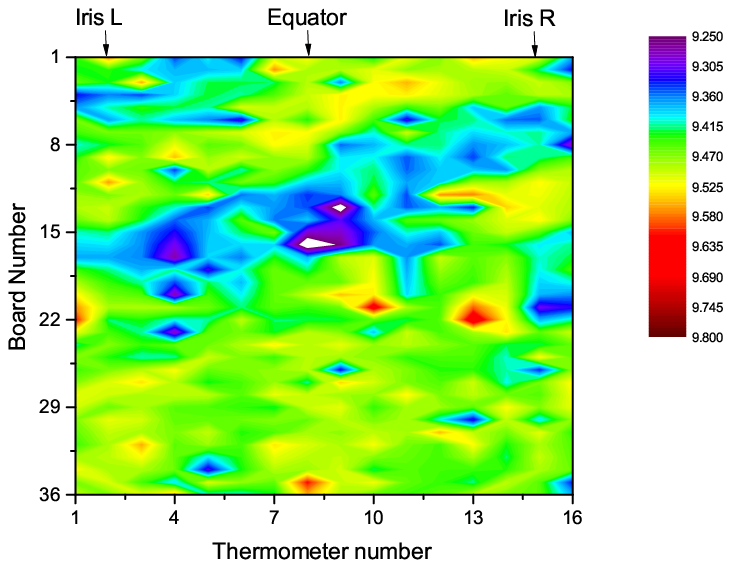}}\\
\subfloat[][]{\includegraphics[]{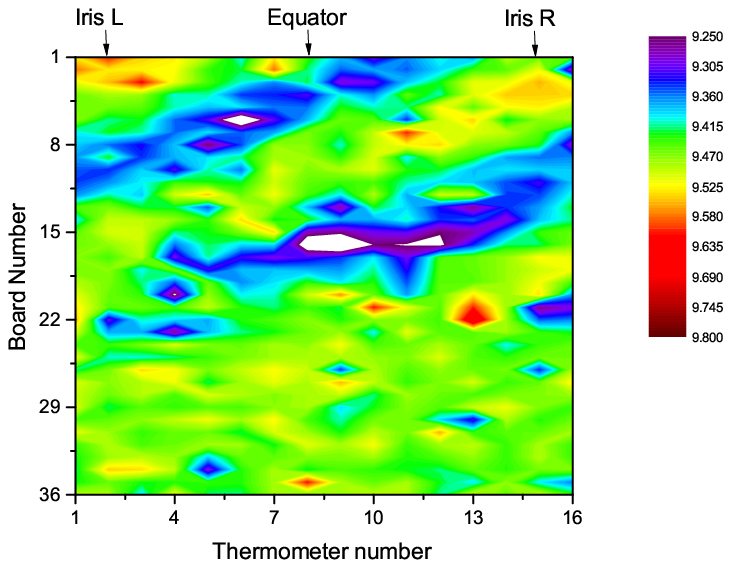}}%
\subfloat[][]{\includegraphics[]{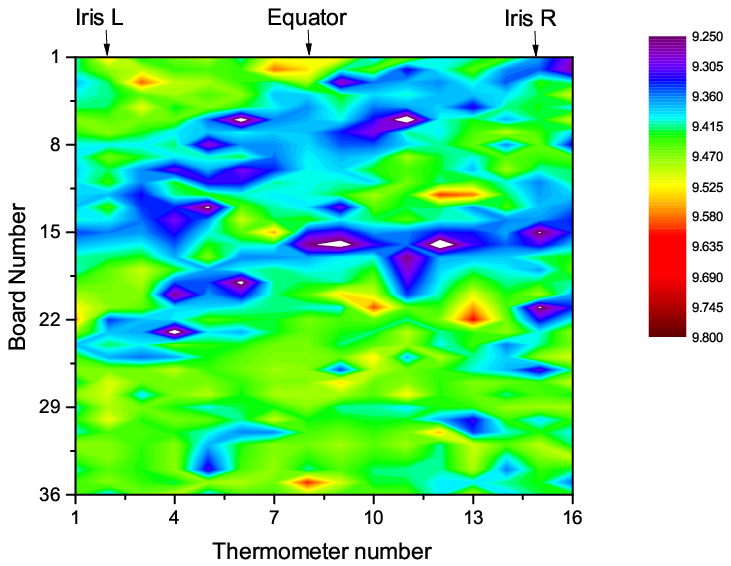}}\\
\subfloat[][]{\includegraphics[]{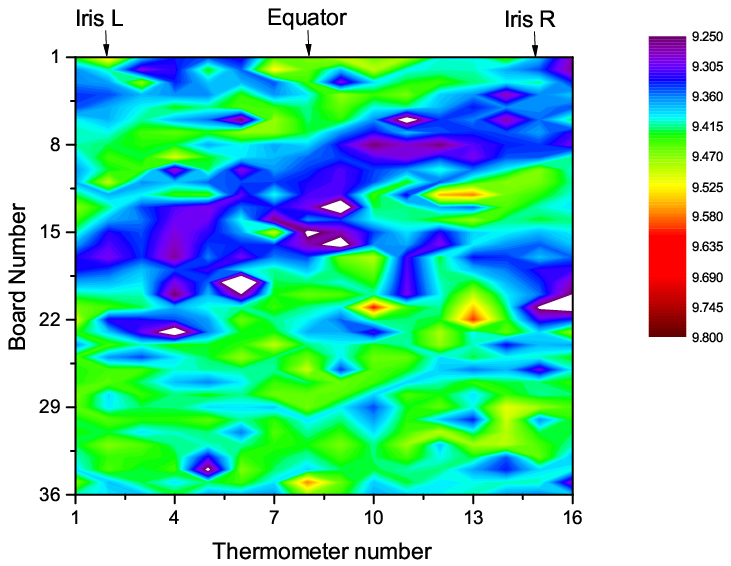}}%
\subfloat[][]{\includegraphics[]{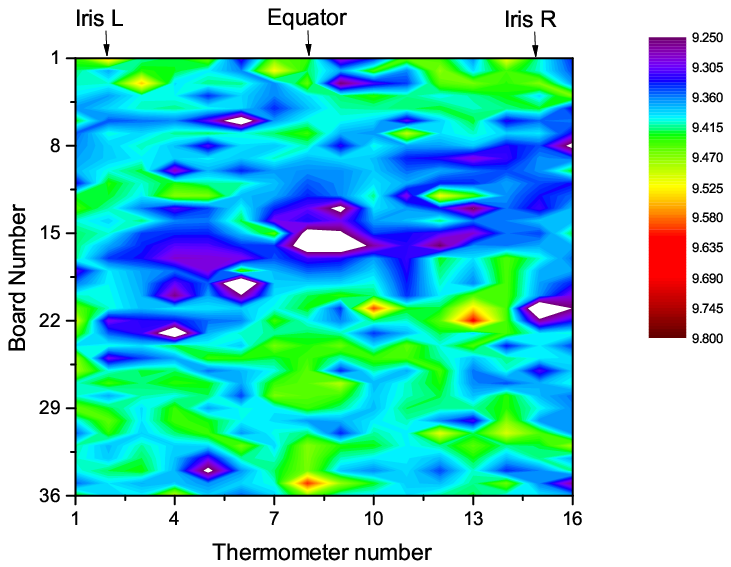}}\\
\caption{}%
\end{figure}
\begin{figure}[H]%
\ContinuedFloat
\centering
\subfloat[][]{\includegraphics[]{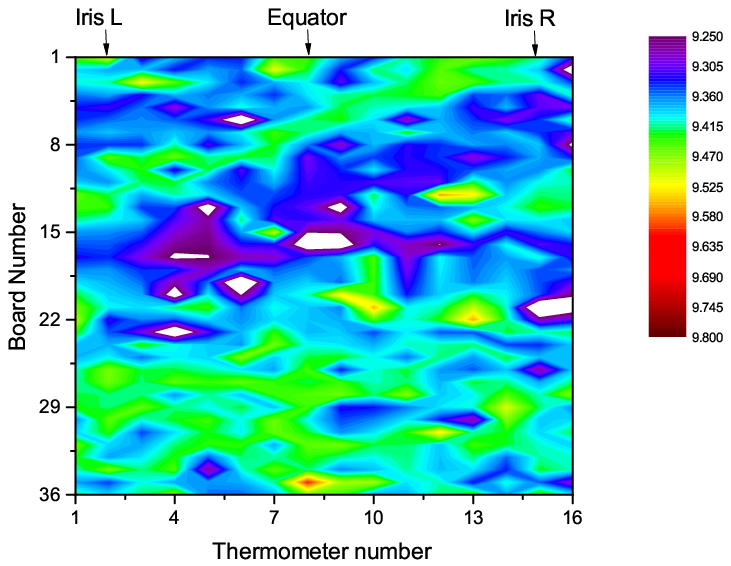}}%
\subfloat[][]{\includegraphics[]{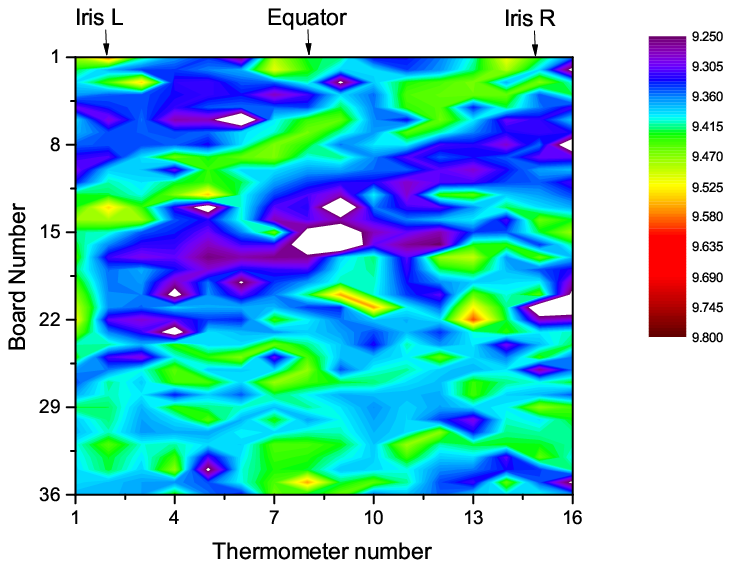}}\\
\subfloat[][]{\includegraphics[]{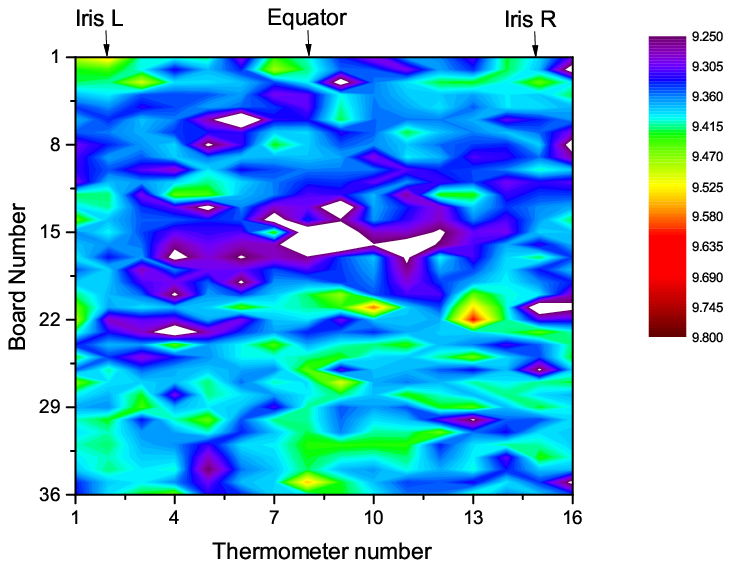}}%
\subfloat[][]{\includegraphics[]{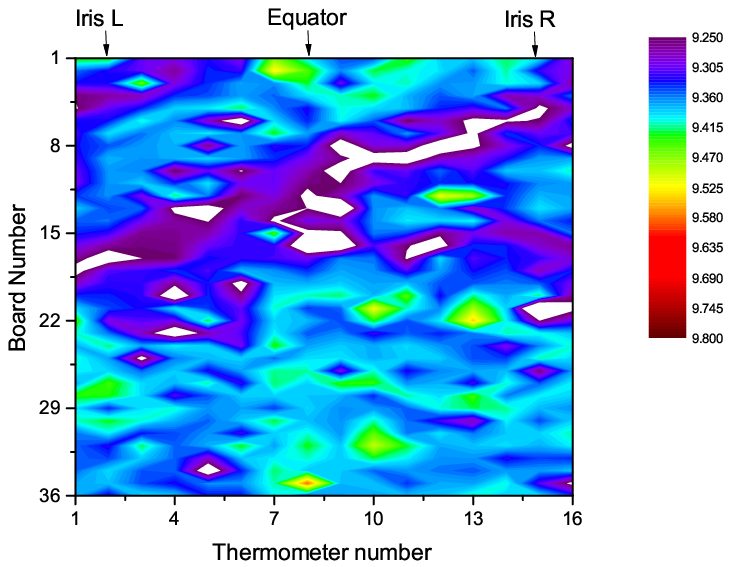}}\\
\subfloat[][]{\includegraphics[]{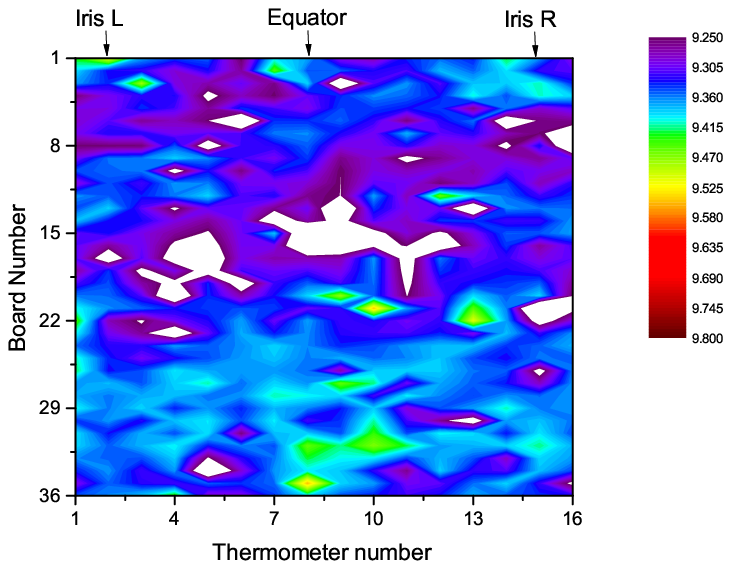}}%
\subfloat[][]{\includegraphics[]{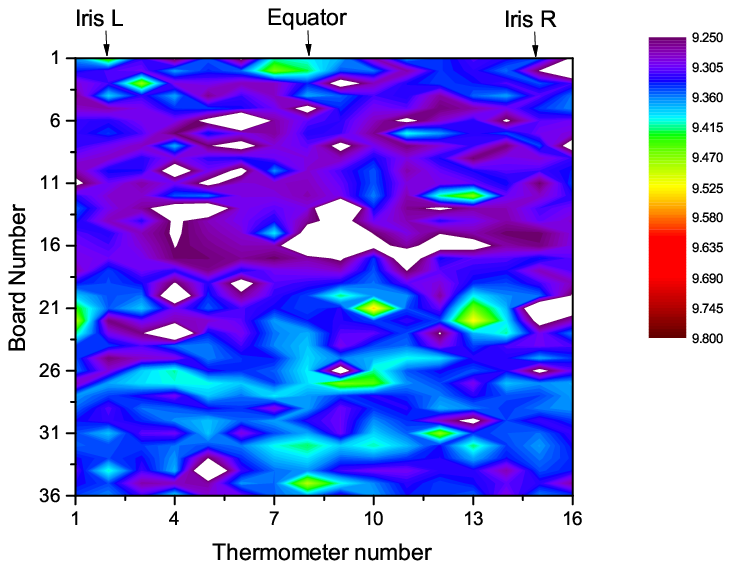}}\\
\caption{}%
\end{figure}
\begin{figure}[H]%
\ContinuedFloat
\centering
\subfloat[][]{\includegraphics[]{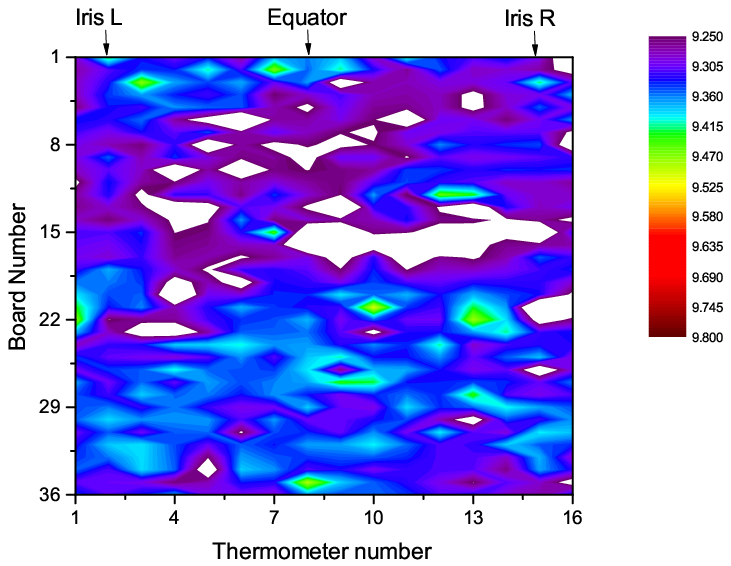}}%
\subfloat[][]{\includegraphics[]{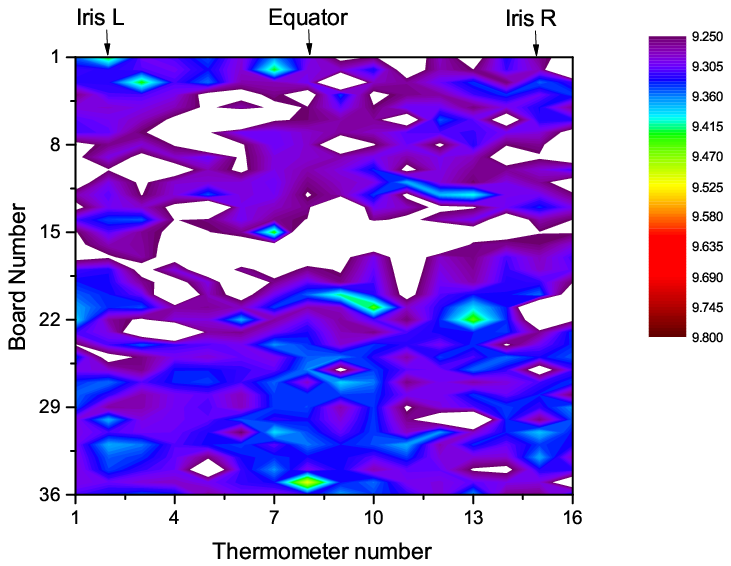}}\\
\subfloat[][]{\includegraphics[]{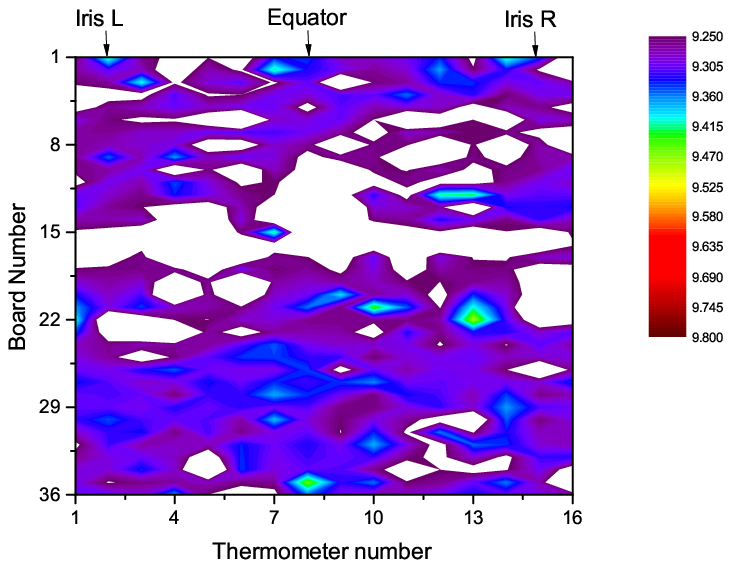}}%
\subfloat[][]{\includegraphics[]{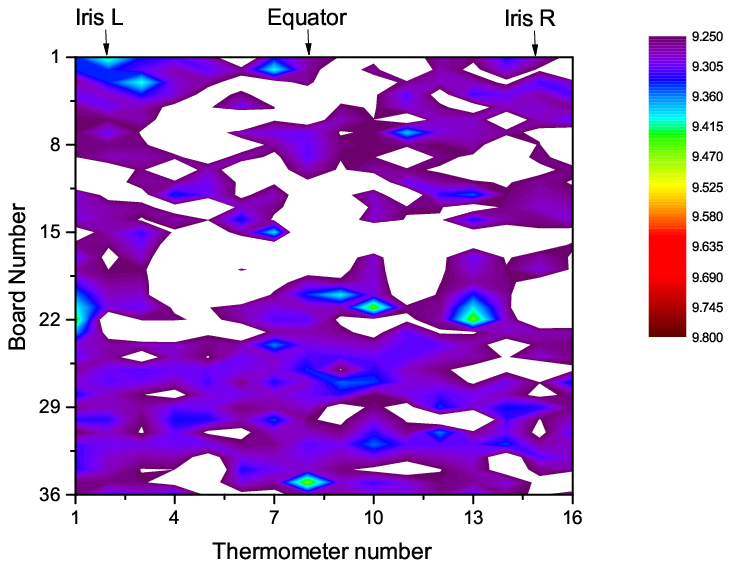}}\\
\subfloat[][]{\includegraphics[]{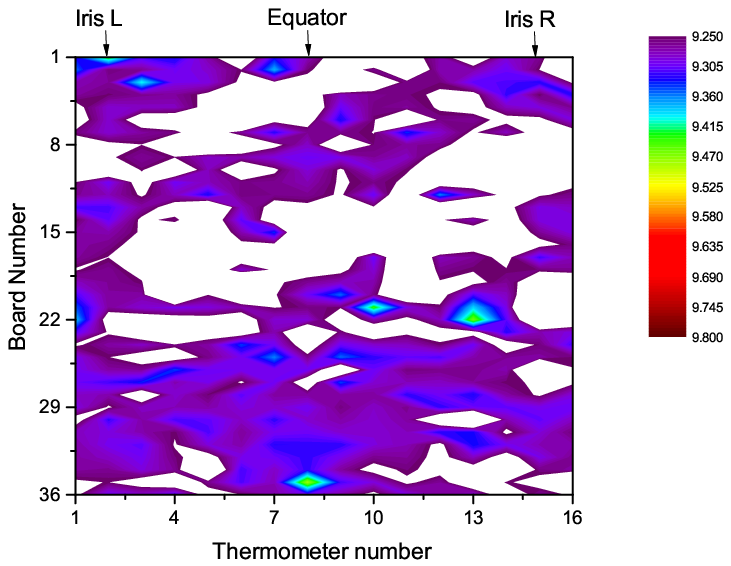}}%
\subfloat[][]{\includegraphics[]{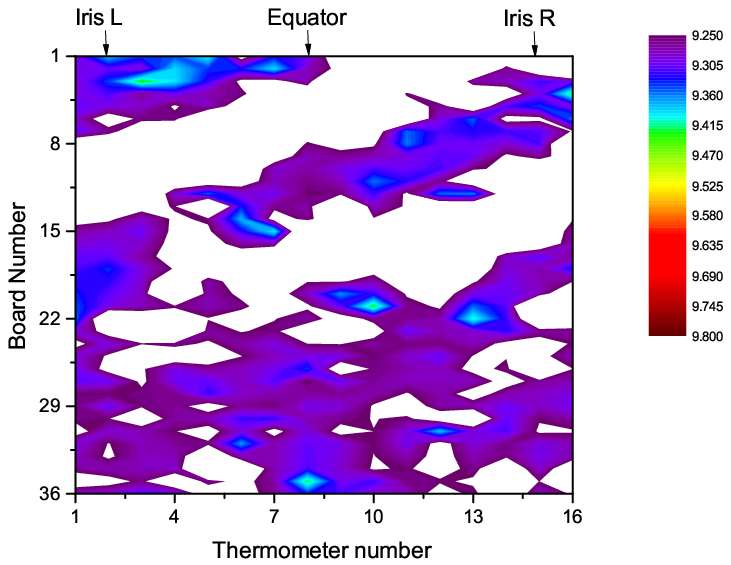}}\\
\caption{}%
\end{figure}
\begin{figure}[H]%
\ContinuedFloat
\centering
\subfloat[][]{\includegraphics[]{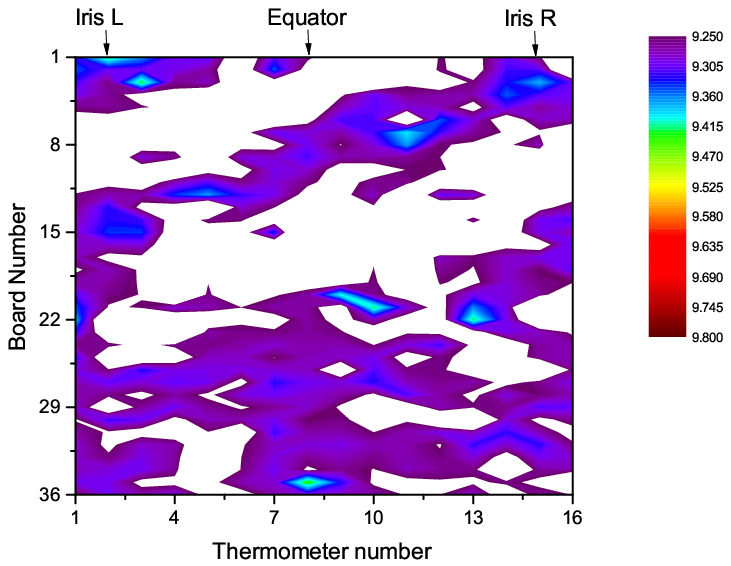}}%
\subfloat[][]{\includegraphics[]{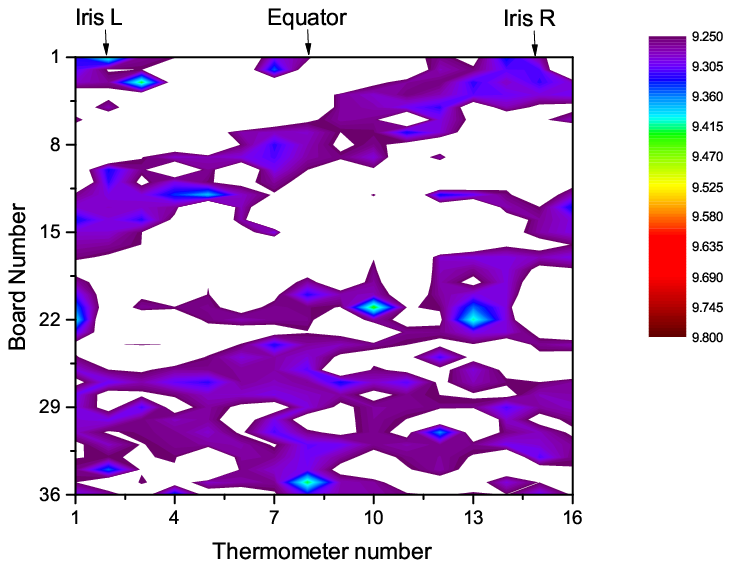}}\\
\subfloat[][]{\includegraphics[]{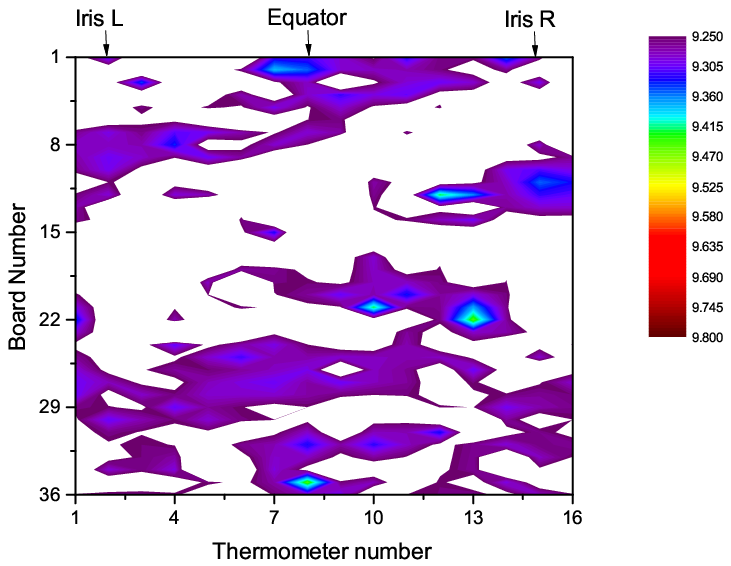}}%
\subfloat[][]{\includegraphics[]{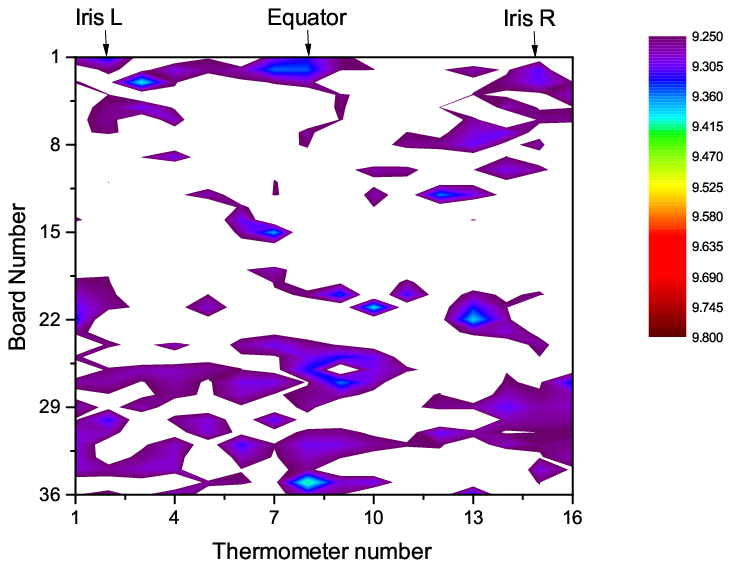}}\\
\subfloat[][]{\includegraphics[]{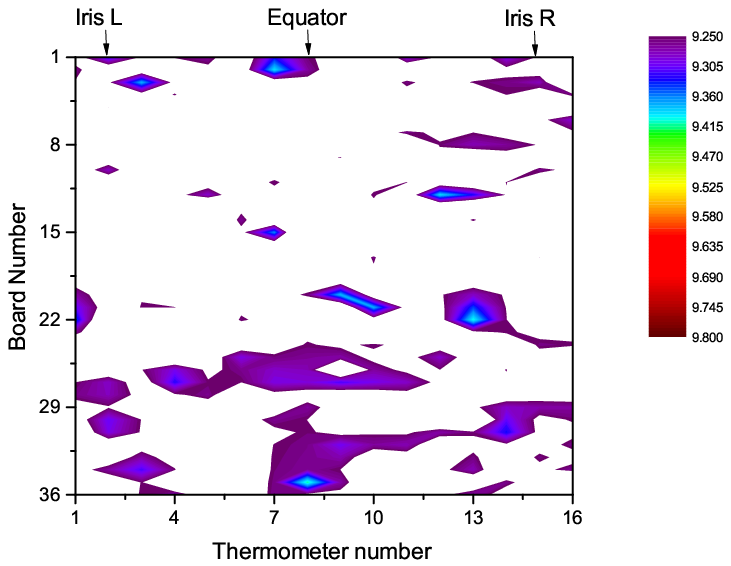}}%
\subfloat[][]{\includegraphics[]{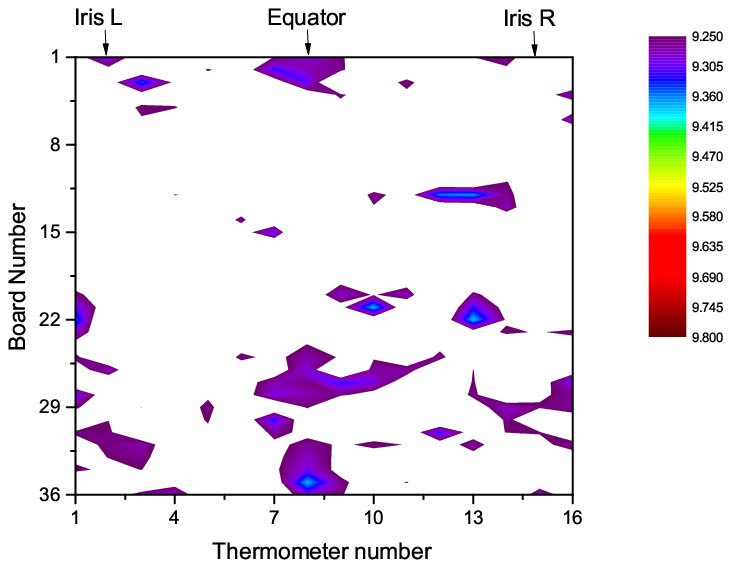}}\\
\caption{}%
\end{figure}
\begin{figure}[H]%
\ContinuedFloat
\centering
\subfloat[][]{\includegraphics[]{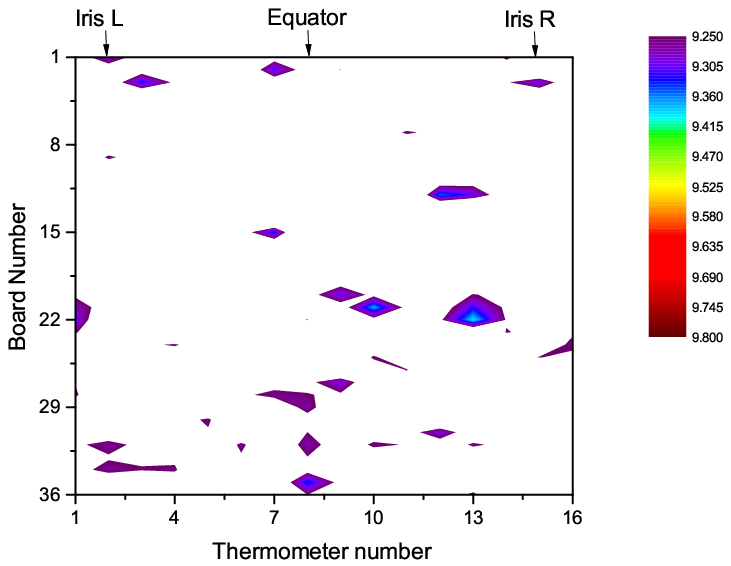}}%
\subfloat[][]{\includegraphics[]{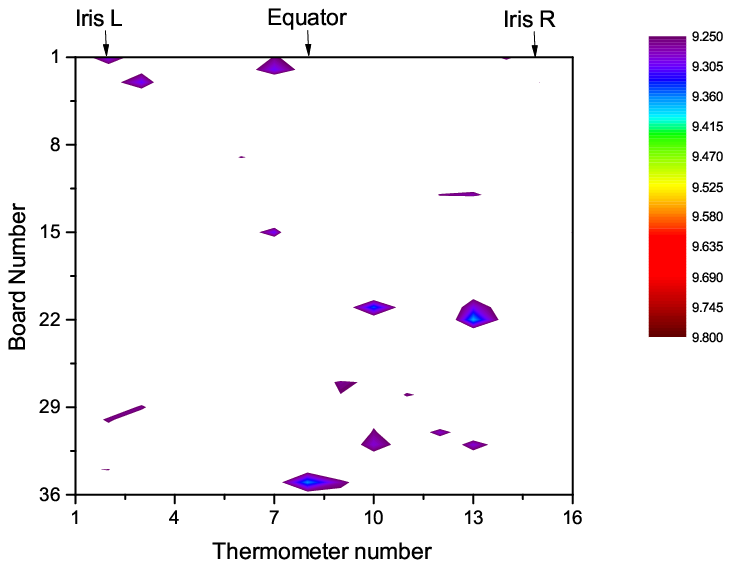}}\\
\subfloat[][]{\includegraphics[]{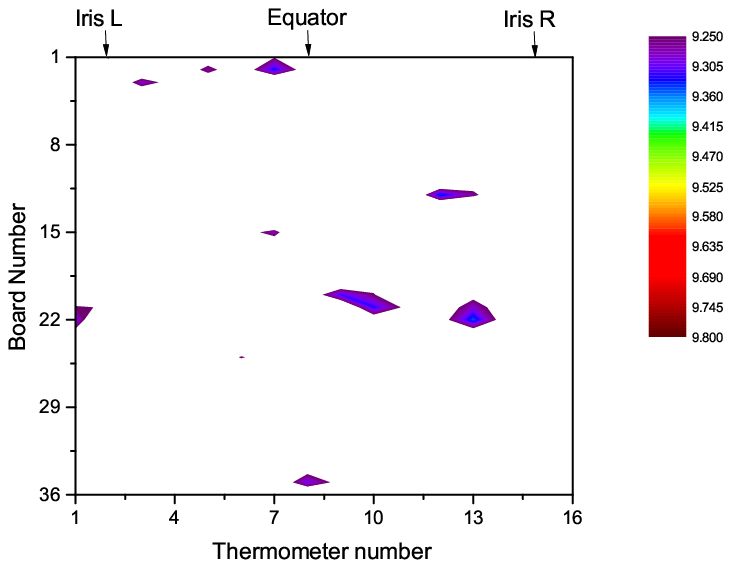}}%
\subfloat[][]{\includegraphics[]{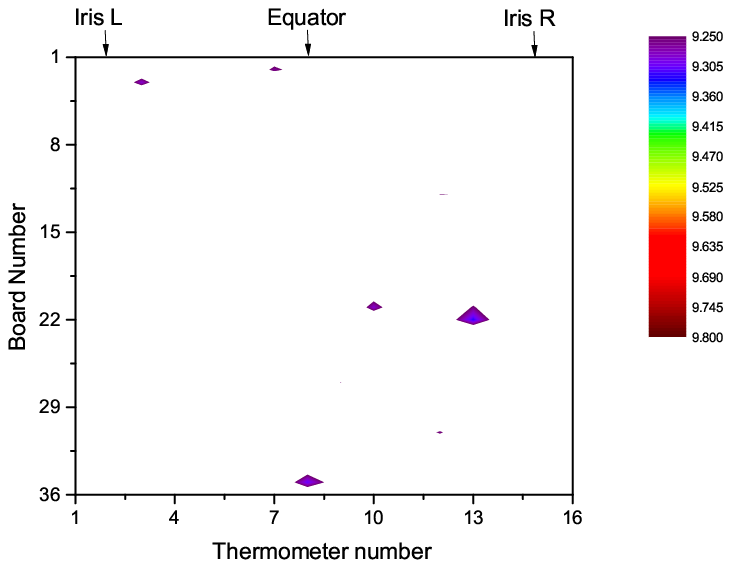}}\\
\subfloat[][]{\includegraphics[]{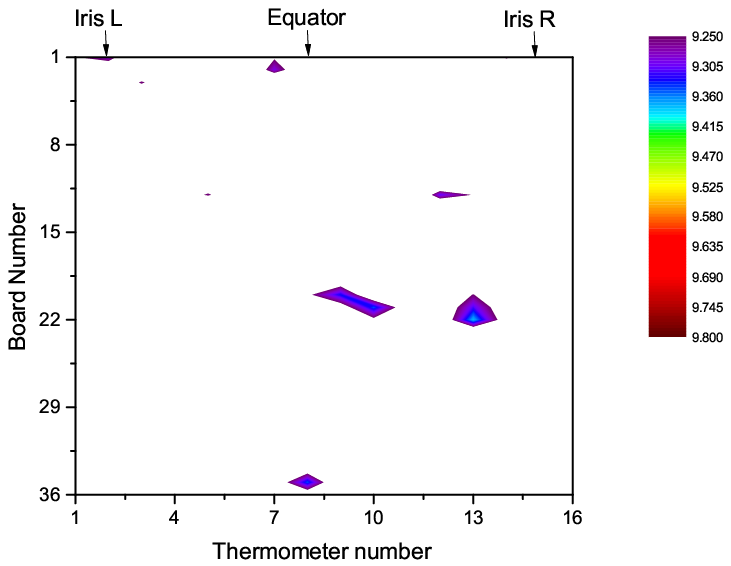}}%
\subfloat[][]{\includegraphics[]{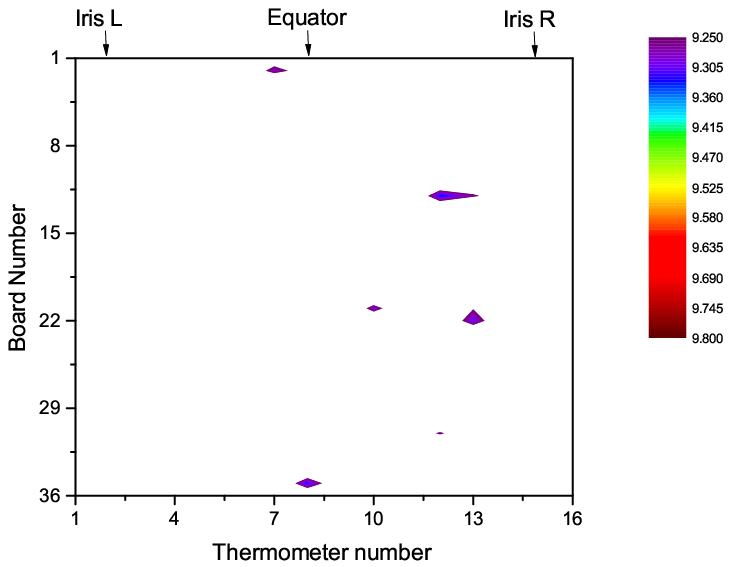}}\\
\caption{}%
\end{figure}
\captionsetup[figure]{labelfont={color=black}}
\begin{figure}[H]%
\ContinuedFloat
\centering
\subfloat[][]{\includegraphics[]{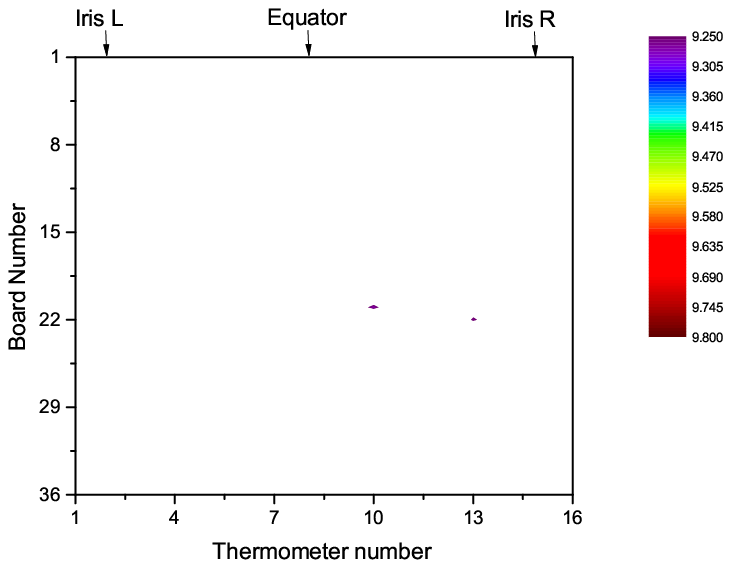}}%
\subfloat[][]{\includegraphics[]{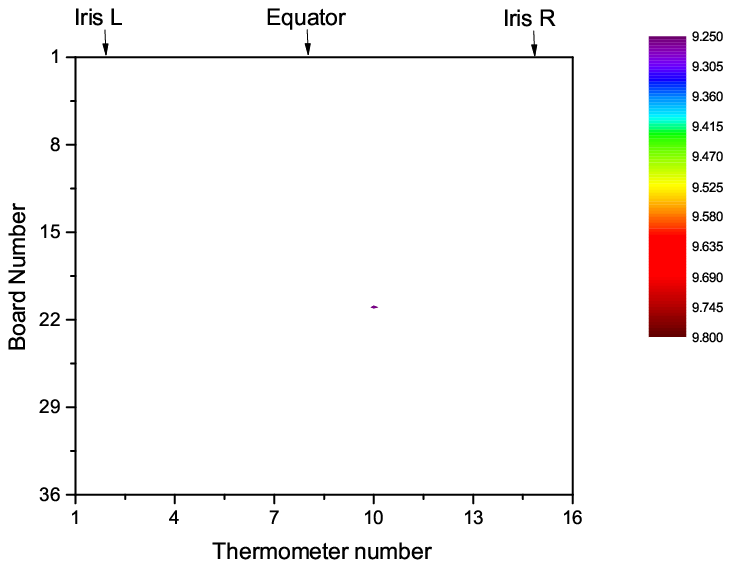}}\\
\subfloat[][]{\includegraphics[]{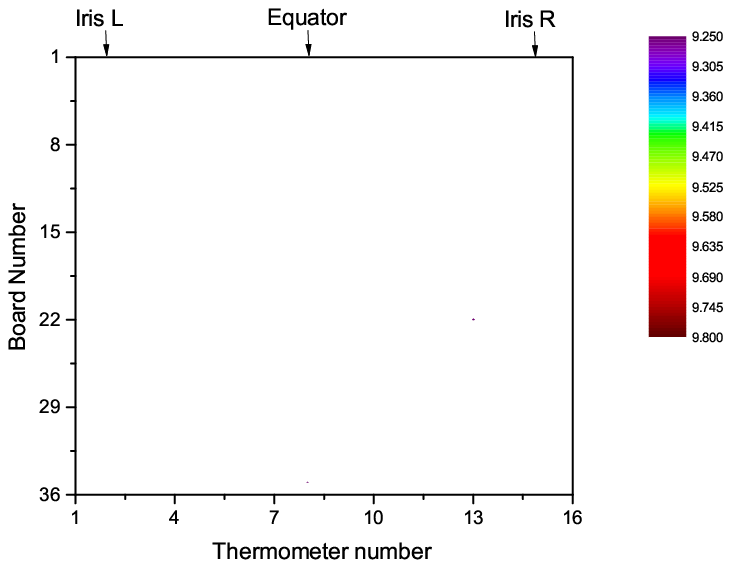}}%
\subfloat[][]{\includegraphics[]{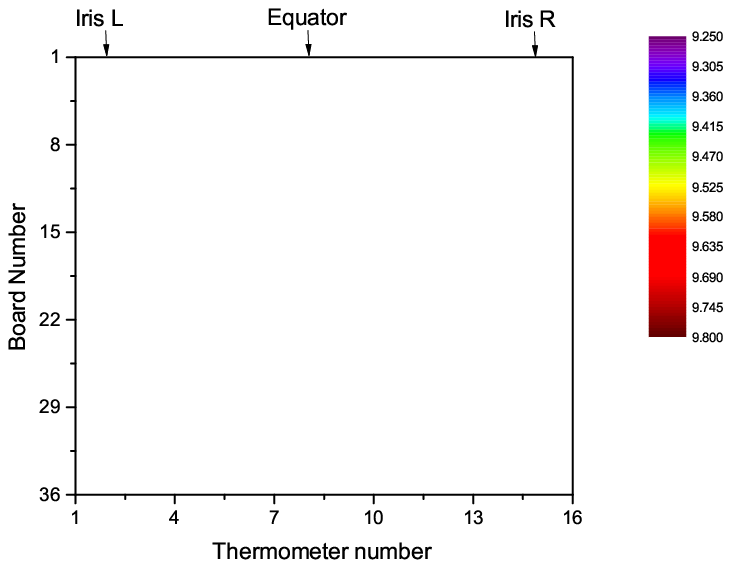}}\\
\caption{Chronological sequence of T-map images acquired during a slow cool-down from 12 K.}%
\label{Slow}
\end{figure}
\indent From the Tmap images appears clear that for the fast cool-down case the superconducting transition proceeds with a sharp SC-NC interface from the bottom to the top of the cavity during the cooling.\\
\indent For the slow cool-down case the superconducting transition occurs with a very different dynamics: at the beginning islands of superconducting phase appear in random positions, oscillating between the SC-NC phases several times before stabilizing below Tc. Then, as the cool-down proceeds, they become larger and larger leading to a situation in which almost all the cavity becomes superconducting but some randomly distributed zones are still normal-conducting.\\
\indent In this situation incomplete Meissner effect easily occurs, the magnetic field expulsion from these NC regions surrounded by the SC phase is not energetically favorable, and magnetic field is randomly trapped all around the cavity.
This scenario clearly prove that the high Q performance shown by nitrogen doped niobium cavities fast cooled from above the critical temperature are due to the sharp SC-NC transition dynamics which optimize the magnetic flux expulsion and minimized the trapped flux dependent surface resistance.
\section*{ACKNOWLEDGEMENTS}
This work was supported by the US Department of Energy, Offices of High Energy Physics and Basic Energy Science, via the LCLS-II High Q Program. Fermilab is operated by Fermi Research Alliance, LLC under Contract No. DE-AC02-07CH11359 with the United States Department of Energy.
%

%
\end{document}